\documentclass[11pt]{article}

\usepackage{graphicx}
\usepackage[margin=1.15in]{geometry}
\usepackage{amsmath}

\newcommand{\ket}[1]{\left| #1 \right>}	%	Ket
\newcommand{\hcm}{\hspace{0.5cm}}	%	Half cm
\newcommand{\qcm}{\hspace{0.25cm}}	%	Quarter cm

\begin{document}

\long\def\symbolfootnote[#1]#2{\begingroup%
\def\thefootnote{\fnsymbol{footnote}}\footnote[#1]{#2}\endgroup}
{\phantom .}
\begin{center}
\textbf {\Large Narratability and Cluster Decomposition} \\
\vspace{1.5em}
{\large Simon Judes}\footnote{{\bf email:} {\it sjudes@gmail.com}} \\
\vspace{0.75em}
Institute for Strings, Cosmology and Astroparticle Physics, Department of Physics\\ Columbia University, New York, NY 10027. \\
\end{center}
\normalsize

\vspace{1ex}
\begin{abstract}
Recently David Albert presented an argument that relativistic quantum theories are \emph{non--narratable}.  That is, specifying the state on every space--like hypersurface in a given foliation of space--time is \emph{not} in general sufficient to determine the states on other hypersurfaces, so the history of the universe cannot be told as a narration of states at successive times.  We show that the system Albert examined to arrive at this conclusion violates cluster decomposition of the S--matrix, a locality requirement satisfied by relativistic quantum field theories.  We formulate the general requirements for a system to display non--narratability, and argue that a large class of systems satisfying them violate the cluster decomposition principle.  
\end{abstract}

\vspace{2ex}

\numberwithin{equation}{section}

\section{Introduction}	\label{intro}
\begin{figure}[h]
\begin{center}
\includegraphics[scale=1.3]{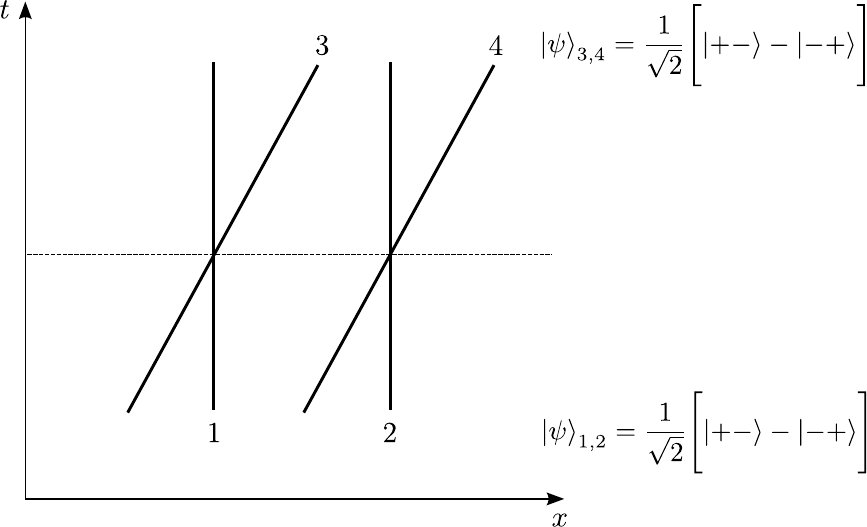}
\end{center}
\caption{Albert's Non--Narratability Example. \label{albert_example}}
\end{figure}

In \cite{AlbertNarratability}, a system of 4 spin 1/2 particles of distinct species was considered, arranged in space--time as in figure \ref{albert_example}.  Particles 1 and 2 are at rest, with 3 and 4 moving toward them with constant velocity, some component of which can be taken to be in (say) the $y$--direction, so that the collisions depicted are the only ones that occur.  And all particles are assumed to be sufficiently massive that it is sensible to speak of them being localized for an arbitrary amount of time.  Since the particles are all distinguishable, there are no permutation symmetry requirements to enforce, but for reasons that will be clear shortly the spin part of the state is nevertheless taken to be a product of two singlet states:
\begin{align*}
 \ket{\psi} = \frac{1}{\sqrt{2}} \Biggl[ \ket{+-}_{12} - \ket{-+}_{12} \Biggr] \frac{1}{\sqrt{2}} \Biggl[ \ket{+-}_{34} - \ket{-+}_{34} \Biggr]
\end{align*}

\begin{figure}[h]
\begin{center}
\includegraphics[scale=1.15]{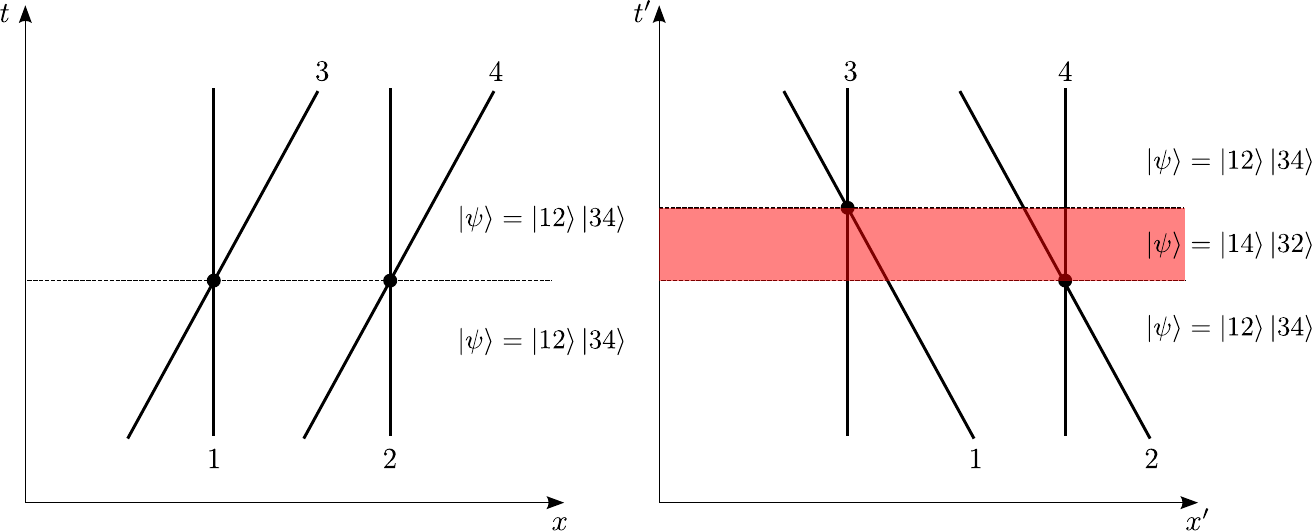}
\end{center}
\caption{Two frames of reference, with the interaction Hamiltonian \eqref{Hint}. \label{albert_example2}}
\end{figure}

The idea is then to consider two possible interaction Hamiltonians, one of which is zero, while the other flips the spins of particles when they collide.  That is, it induces the following unitary transformation on the spin Hilbert space:
\begin{align}
	\ket{++} &\qcm\longrightarrow\qcm \ket{++}, \label{Hint}
	&\ket{+-} \qcm\longrightarrow\qcm \ket{-+} \\ \nonumber
	\ket{-+} &\qcm\longrightarrow\qcm \ket{+-}, 
	&\ket{--} \qcm\longrightarrow\qcm \ket{--} 
\end{align}
There are two such collisions in figure \ref{albert_example}.  Denoting the singlet state of particles $a$ and $b$ as $\ket{ab}$, we have:
\begin{align}\label{evolution}
	\ket{12}\ket{34} \xrightarrow{\text{1 hits 3}} \ket{14}\ket{32} \xrightarrow{\text{2 hits 4}} \ket{12}\ket{34}
\end{align}
But since the collisions happen simultaneously, the middle step in \eqref{evolution} is bypassed, and the transformation has no effect on the evolution of the spin state. Thus the state history is the same no matter whether the interaction Hamiltonian vanishes or is given by \eqref{Hint}.  On the other hand, in almost any other frame of reference the collisions are not simultaneous,\footnote{Simultaneity is preserved if the boost has no component in the $x$--direction.} as shown in figure \ref{albert_example2}.

In the primed frame of reference, the state history \emph{does} depend on the choice of Hamiltonian.  Without interactions, the evolution of the spin state is the same as in the original frame.  But with the interactions, there is a period between the two collisions where the spin state is $\ket{14}\ket{32}$ rather than $\ket{12}\ket{34}$.\footnote{A point passed over here is that the Lorentz transformation of the spin state is in general non--trivial, even when there are no interactions.  The spin state transforms by an element of the little group, which for a massive particle in 3+1 dimensions is $SO(3)$, the 3d rotation group.  In the example above the state is a product of two singlets, which has zero angular momentum, so the transformation is indeed trivial if the Hamiltonian is free.  }

So a complete specification of the state--history in one frame of reference is insufficient to determine the state--history in any other frame of reference.  Albert concludes that according to relativistic quantum mechanics, the history of the universe cannot be told as a narrative --- there is more going on than a state description at each time: 
\begin{quotation}
``In order to present the entirety of what there is to say about a Relativistic Quantum-Mechanical world, we need to specify, separately, the Quantum-Mechanical state of the world associated with every\footnote{Presumably one doesn't need the state on \emph{every} space--like hypersurface, but only on enough foliations so that any two events will fail to be simultaneous on at least one of them.} separate space--like hypersurface.'' \cite{AlbertNarratability}
\end{quotation}
Though the dependence of Lorentz transformations on the Hamiltonian has been known for a long time,\cite{WeinbergQFT1}\cite{Fleming1} the appearance of non--narratability is surprising, and its consequences have been subject to some debate.\cite{0907.5294v1}  

But it is far from clear that the argument above is sufficient to put the narratability of the universe in question.  The Hamiltonians of relativistic quantum field theories satisfy special constraints: the corresponding S--matrices are Lorentz invariant, and obey the cluster decomposition principle.  We argue here that an interaction of the kind required to flip the spins as in figure \ref{albert_example} cannot arise in a local quantum field theory like the standard model, because the S--matrix would not satisfy cluster decomposition.\footnote{A variant of Albert's system with only two particles is considered in \cite{0907.5294v1}, but with a manifestly nonlocal interaction.}  

This leaves open the question of whether relativistic quantum field theories are narratable.  It might be that a system analogous to Albert's can be found that is consistent with Lorentz invariance and the cluster decomposition principle.  We propose sufficient conditions for a system to demonstrate non--narratability, and a show that a class of such systems of which Albert's is a member are inconsistent with cluster decomposition of the S--matrix, and therefore cannot arise from a relativistic quantum field theory.

\section{Cluster Decomposition}
A position space S--matrix is said to obey cluster decomposition when its connected part\footnote{The connected part of the S--matrix is the part that corresponds to all incoming and outgoing particles interacting with one another --- its nonzero contributions come from topologically connected Feynman diagrams.} vanishes if any of the incoming or outgoing particles are moved very far away from the others.  This is a locality constraint --- it amounts to the demand that scattering amplitudes don't depend on facts about distant parts of the universe.  In momentum space cluster decomposition requires that the connected part of the S--matrix contain a single momentum conservation delta function, and no additional delta functions containing only proper subsets of the momenta.  This turns out to be equivalent to demanding that the interaction Hamiltonian also have only an overall momentum conservation delta function.\cite{WeinbergQFT1}

In \cite{WeinbergQFT1} cluster decomposition is regarded as an axiom, which along with Lorentz invariance of the S--matrix, leads naturally (if not inevitably) to quantum field theory.  Alternatively one could decide for empirical reasons to look only at S--matrices arising from local quantum field theories, and observe that they all obey cluster decomposition.  The important point is that cluster decomposition is a property of our best current approximation to the laws of nature.

It's useful to distinguish the kind of nonlocality this prohibits, from the kind that it still allows.  Consider for example Bohm's version of the EPR experiment.  Here there are two distant observers measuring electrons in the singlet state.  Their results are always anticorrelated, and as Bell showed there must be a non--local influence to account for these correlations, but there is no possibility of sending a signal from one observer to the other because the statistics of the measurements carried out by any one observer are unaffected by what the other one does.  If the actions of one observer could affect the \emph{probabilities} observed by the other observer then a signal could easily be sent.  This is the kind of behavior the cluster decomposition principle prohibits.

\subsection{Albert's Example}
\begin{figure}[t]
\begin{center}
\includegraphics[scale=1]{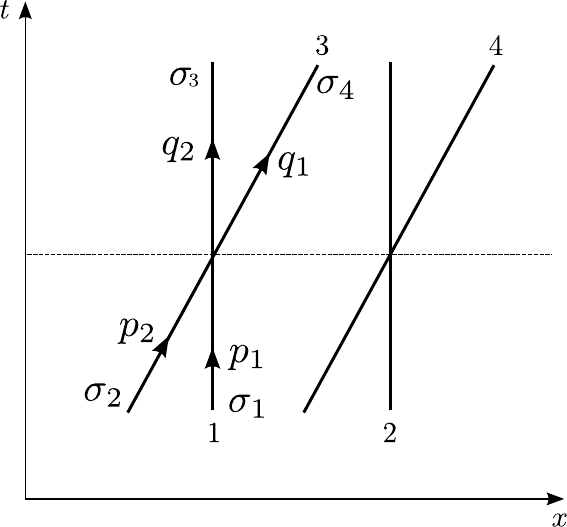}
\end{center}
\caption{Momentum and spin labels. \label{albert_example_momenta}}
\end{figure}

We now attempt to formulate the interactions in Albert's example in a Fock space formalism.  We have a separate sector for each of the 4 species of particle, and the essentially unique interaction term corresponding to the spin flip is:\footnote{In principle \eqref{spin_flip_interaction} could contain additional smooth functions of the momenta, such as the factors of $E_{\vec{p}_i}^{-1/2}$ required to make the interaction Lorentz invariant.  For our purposes these don't matter too much because they don't affect the additional delta function that violates cluster decomposition. }
\begin{align} 
	V = \sum_{(\text{pairs }a,b)}\sum_{\sigma_1\sigma_2\sigma_3\sigma_4}\int{\rm d}^3p_1&\int{\rm d}^3p_2\int{\rm d}^3p_3\int{\rm d}^3p_4  \nonumber \\ \label{spin_flip_interaction}
	& \cdot a^{\dag}(q_2, \sigma_3) b^{\dag}(q_1, \sigma_4) a(p_1, \sigma_1) b(p_2, \sigma_2)  \\
	& \cdot \delta_{\sigma_3\sigma_2}\delta_{\sigma_4\sigma_1} \delta^3(\vec{q}_2 - \vec{p}_1)\delta^3(\vec{q}_1 - \vec{p}_2) \nonumber
\end{align}
Here $a$ and $ b$ are the annihilation operators for two of the four species of fermions, and the momentum and spin labels are defined in figure \ref{albert_example_momenta}.   The outermost sum is over all pairs of particle species, so that the collision of any two produces the interaction.  And the delta functions are necessary to ensure that the spins flip while the momenta are unchanged, as they must be if the state history is to be consistent with the free Hamiltonian in the frame where the collisions are simultaneous. 

The momentum delta functions can be rewritten: $\delta^3(\vec{q}_1 + \vec{q}_2 - \vec{p}_1 - \vec{p}_2)\delta^3(\vec{q}_1 - \vec{p}_2)$.  And this is the problem.  In addition to the overall conservation of momentum delta function there is another, enforcing a stricter kinematic requirement --- that the momenta are unchanged by the interaction.  This extra delta function is precisely what is forbidden by cluster decomposition.

The problem is that to demand that the interaction be local, and that it leave the momenta unchanged is overconstraining.  One can define the interaction on momentum eigenstates, or on position eigenstates, but since they are linearly related, not in general on both.  And what the violation of cluster decomposition shows is that once the interaction is defined so as to leave the momenta fixed, locality is out of the question.  One can have small momentum transfers, but only at the expense of non--local spin interactions; or one can localize the interaction, but large momentum transfer must follow.  
 
\subsection{More General Examples}
Is the violation of cluster decomposition a peculiarity of this particular case?  Or will any interaction Hamiltonian suitable for demonstrating non--narratability also violate cluster decomposition?  Given that cluster decomposition is true of the standard model, it seems important to find a version of Albert's argument consistent with it.  Whether such a model exists surely bears on whether \emph{our} universe is narratable, even if more general relativistic quantum systems are not.

It is helpful to examine the Hamiltonian dependence of Lorentz transformations from the point of view of the Poincar\'{e} algebra, whose generators: $\vec{P}, H, \vec{J}, \vec{K}$  enact space--time translations, rotations, and boosts respectively.  If $H_0, \vec{P}_0, \vec{J}_0, \vec{K}_0$ are the generators on non--interacting particles, then the effect of adding interactions is usually taken to be:
\begin{align*}
 	H = H_0 + V \hcm \vec{P}=\vec{P}_0 \hcm \vec{J}=\vec{J}_0
\end{align*}
But the commutation relation $[P_i,K_j] = i\delta_{ij}H$ prevents the imposition of $\vec{K}=\vec{K}_0$. Writing instead $\vec{K}=\vec{K}_0+\vec{W}$, we find: $[\vec{K}_0, V] = -[\vec{W}, H]$, from which the matrix elements of $\vec{W}$ between energy eigenstates can be defined. 

What is needed for an example like that in figure \ref{albert_example2} is a state history $\ket{\psi_t}$ and an interaction Hamiltonian $V$ satisfying:
\begin{align}
	e^{i(H_0+V)t} \ket{\psi_0} &= c e^{iH_0t} \ket{\psi_0} \label{req1} \\
	W_i \ket{\psi_t} &\neq c' \ket{\psi_t} \label{req2}
\end{align}
where $c$ and $c'$ are complex constants than can in principle depend on time.  Equation \eqref{req1} ensures that $\ket{\psi_t}$ is a solution of the Schr\"{o}dinger equation with either $H_0$ or $H_0+V$ as the Hamiltonian, while \eqref{req2} demands that the Lorentz transformation depends nontrivially on the Hamiltonian.  

To satisfy \eqref{req1}, we need an interaction Hamiltonian that leaves the space--time trajectories of the particles unchanged.  But any interaction term of degree 3 or higher in the creation and annihilation operators will contain additional momenta that will have to be constrained by delta function factors to ensure that no space--time scattering takes place, replicating the state history of the free Hamiltonian.  Thus no such interaction can be compatible with cluster decomposition.

If no example like that in figure \ref{albert_example2} could be constructed that was consistent with cluster decomposition and gave a Lorentz invariant S--matrix, one might conclude that although some possible quantum mechanical worlds are not narratable, our world in fact \emph{is} narratable, because a Lorentz invariant S--matrix satisfying cluster decomposition is a consequence of quantum field theories. 

But the above argument is not sufficient to reach this conclusion, because the conditions \eqref{req1} and \eqref{req2} are somewhat too strong --- we can look for a state history consistent with \emph{two} interacting Hamiltonians rather than one interacting and one free:
\begin{align} 
	e^{i(H_0+V_1)t} \ket{\psi_0} &= c e^{i(H_0+V_2)t} \ket{\psi_0} 
\end{align}
and the analogue to $W_i$ is now the difference between the two interacting boost generators.  Equation \eqref{req1} then corresponds to the special case: $V_2=0$.

\section{Conclusion}
The system presented in \cite{AlbertNarratability} is not narratable, so it is natural to ask whether more realistic theories also fail to be narratable.  We formulated the general requirements for a theory to be non--narratable, and showed that cluster decomposition --- a locality constraint true of quantum field theories --- rules out the class of examples where one of the Hamiltonians is the free Hamiltonian $H_0$.  One possibility still remains for a demonstration of non--narratability: to replace $H_0$ with some other Hamiltonian $H_1$.  Ultimately the most interesting question is whether \emph{our} universe is narratable, so we would like $H_1$ to be the Hamiltonian of our best current theory of particle physics.  The existence of another Hamiltonian satisfying Lorentz invariance and cluster decomposition, and consistent with experiments performed up to now, yet differing in its predictions of experiments performed in highly boosted frames would be a remarkable discovery.

\subsection*{Acknowledgments}
\ \ \ \ \
The author thanks Lucien Hardy, Joshua Rosaler, and especially David Albert for useful discussions.

%\bibliographystyle{unsrt}
%\bibliography{philphys}

\end{document}